\documentclass{PoS}

\def\apj{ApJ}                 % Astrophysical Journal
\def\apjl{ApJ}                % Astrophysical Journal, Letters
\def\aap{A\&A}                % Astronomy and Astrophysics
\def\mnras{MNRAS}             % Monthly Notices of the RAS
\def\nat{Nature}              % Nature
\usepackage{lineno}
\title{Search for very-high-energy emission with HAWC from GW170817 event}

\ShortTitle{Follow up to GRB 170817A}

\author{\speaker{Antonio Galv\'an-G\'amez} ; \ Nissim Fraija and M. Magdalena Gonz\'alez \\
The HAWC Collaboration\footnote{For collaboration list, see PoS(ICRC2019) 1177.}\\
{\itshape \href{https://www.hawc-observatory.org/collaboration/icrc2019.php}{https://www.hawc-observatory.org/collaboration/icrc2019.php}}\\
        Instituto de Astronom\'ia, UNAM.\\
        E-mail: \email{agalvan@astro.unam.mx}, \email{nifraija@astro.unam.mx}, \email{magda@astro.unam.mx}}

\abstract{The detection of the gravitational wave GW170817 defined a breakthrough in multi-messenger astronomy. For the first time, a gravitational wave transient detected by the Laser Interferometer Gravitational-Wave Observatory (LIGO) and Virgo interferometer was associated with a faint electromagnetic gamma-ray counterpart reported by the Gamma-ray Burst Monitor (GBM) aboard on the Fermi satellite. GRB 170817A was followed up by an enormous observational campaign covering a large fraction of the electromagnetic spectrum. In this work, we use the data from High Altitude Water Cherenkov (HAWC) gamma-ray observatory to search for very-high-energy (VHE) TeV photons in coincidence with the X-ray emission from GRB 170817A. Since no counts were observed up to $\sim$120 days after the trigger time, we derive and report the corresponding upper limits in the energy range from 1 to 100 TeV.  In addition, we extend the analysis to GRBs with similar features proposed by A. von Kienlin.}
\FullConference{36th International Cosmic Ray Conference -ICRC2019-\\
		July 24th - August 1st, 2019\\
		Madison, WI, U.S.A.}

\begin{document}

\section{Introduction}
One of the events defining a breakthrough in the multi-messenger astronomy is the detection of the gravitational wave (therefore GW) GW170817 \cite{GCNGW...LW,GCNGW...LW2} by the LIGO and Virgo observatories. This gravitational wave was identified as a binary system consisting of neutron stars. The merger was associated with a short Gamma Ray Burst (from now sGRB or GRB) produced $\sim 2$s \cite{2017ApJ...848L..12A} after the GW event. The GRB was detected by the Gamma-ray Burst Monitor (GBM) on board the Fermi spacecraft \cite{2017GCN.21520....1V}. Since the detection of the prompt emission, several of optical telescopes on ground started one of the most exhaustive follow up campaigns of the modern astronomy. The first detection of the optical counterparts came from the Large Binocular Telescope Observatory (LBT) \cite{2017GCN.21520....2V} $\sim$10 hours after the merger. The most interesting detection to the purposes of TeV counterpart searches came from the observation of the X-ray \cite{2017Natur.551...71T,2017ApJ...848L..20M} and radio \cite{2017Sci...358.1579H} emissions 9 and 16 days after the prompt emission, respectively. For the first time in GRBs astronomy, the maximum fluxes were reached $\sim$ 120 days after the trigger time. Different models based on external shocks  have been developed to explain this burst \cite{2017arXiv171111573M, 2017arXiv171203237L, 2019ApJ...871..200F, 2019ApJ...871..123F, 2019arXiv190600502F, 2019arXiv190407732F}. In general, TeV photons are likely generated by synchrotron self-Compton (SSC) emission from X-ray photons \cite{2019arXiv190513572F,2019arXiv190706675F,2015ApJ...804..105F, 2016ApJ...818..190F,2017ApJ...848...15F,2014MNRAS.437.2187F, 2019ApJ...879L..26F}. For this reason, in this work we search for TeV emission up to the time of the maximum X-ray flux.

Recently von Kienlin et. al. \cite{2019arXiv190106158V}, using the data from the Fermi-GBM 10 yr catalog\footnote{FERMIGBRST - Fermi GBM Burst Catalog}, identified, applying a two-step selection supported by a statistical criteria using bayesian block analysis, 13 sGRBs  with similar features to GRB 170817A. Since HAWC scans the whole sky continuously, it is ideal and possible to search for TeV emission in coincidence with these bursts.

The paper is arranged as follows. In Section \ref{Ch:FollowUp} we present the bursts studied here. In Section \ref{Ch:Analysis} we describe briefly the analysis performed. Finally, in Section \ref{Ch:Conclusions}, discussion and results are presented.

%Finally, in Section \ref{Ch:Conclusions} we give a brief summary.

\section{Follow-up to GRB 170817A and sGRBs alike.} \label{Ch:FollowUp}
GRB 170817A triggered Fermi GBM at 2017 September 17 12:41:20 UTC. The Fermi GBM localization constrained this burst to a region at $\alpha=12^{\rm h}20^{\rm m}$ and $\delta=-30^\circ$ (J2000.0). Immediately, GRB 170817A was followed up by a massive observational campaign covering a large fraction of the electromagnetic spectrum.  
In X-ray bands, this burst was detected by the Chandra and XMM-Newton satellites, in optical bands, non-thermal observations was revealed by the Hubble Space Telescope and in radio, bands at 3 and 6 GHz were identified by the Very Large Array (VLA).

The GW170917/GRB 170817A event a few hours later entered the field of view of two of the TeV $\gamma$-ray observatories: the High Energy Stereoscopic System  (H.E.S.S.) Telescope and the HAWC observatory.   Observations with the HAWC experiment started on 2017 August 17 at 20:53 UTC and finished 2.03 hr later. %These authors used a power-law spectrum with a temporal decay index of $-2.5$. 
Although no significant excess of counts was detected by HAWC, an upper limit of $1.7\times 10^{-10}\,{\rm erg\, cm^{-2}\,s^{-1}}$ for an energy range of 4 - 100 TeV was derived.

Recently, A. von Kienlin in \cite{2019arXiv190106158V} analyzed the GBM catalog looking for similarities with other sGRBs. These authors proposed a sample of sGRBs with similar characteristics to GRB 170817A. %: has proposed a sample of sGRB with similars to the GRB170817A. 
This is derived from the GBM data by applying cuts on the observable features (e.g.  $T_{90}$ < 5 s, the main pulse was followed by a soft tail emission, etc). Those bursts are clearly interesting to search for TeV emission at long time scales as in GRB 170817A. 

\section{Analysis} \label{Ch:Analysis}

The HAWC observatory is located next to the volcano Sierra Negra in the state of Puebla in Mexico at an altitude of 4,100m a.s.l. HAWC is constituted by 300 Water Cherenkov Detectors (WCDs) as the main array and 345 smaller detectors to improve its sensitivity to showers of the highest energies. HAWC has an instantaneous field of view of 2 sr, a duty cycle > 95$\%$ and an angular resolution $\gtrsim$ 0.1$^{\circ}$. The optimal sensitivity of HAWC lies in the range of declination of $\delta \in [-26^{\circ}, 64^{\circ}]$ and energies of 1 and 100 TeV. 

The details on the adopted analysis in this work, such as event reconstruction, systematic uncertainties, background subtraction and map making, are described in \cite{2017ApJ...841..100A}. Here, specific details relevant to our analysis are presented. 

Although HAWC continuously records extensive air showers in all directions over the horizon, its effective area is a function of the zenith angle of the primary gamma-ray. It is optimal for values < 45$^{\circ}$ and the contribution from those events outside a cone with open angle of $\sim$45$^{\circ}$ is small. As shown by   \cite{2017ApJ...841..100A} for a source like the crab nebula, it is expected that 90$\%$ of the signal arrives within the $\sim$4 central hours of the transit. The detection efficiency is not uniform and depends on the source declination. On the other hand, the energy sensitivity is a function of the declination of the source and the spectral index of the source as the Figure 10 from \cite{2017ApJ...843...40A} shows.

Sidereal day sky maps divided into 9 bins as a function of the multiplicity of the shower events are constructed as described in \cite{2017ApJ...841..100A}. The maps are combined to cover consecutive time intervals beginning at the GW trigger. For the case of GRB 1701817A we have considered 10 consecutive time intervals in logarithmic scale covering a total of 120 days. For other bursts we took time intervals of 1, 10 and 100 days. The background fluctuations are well approximated by a Poisson distribution thus, the significance distribution is well described by a normal function \cite{2012ApJ...750...63A}. Upper limits of the flux are derived assuming a spectral index of $-2.5$ with a pivot energy of 1 TeV on the GRBs location using a Feldman-Cousins confidence interval approach and considering the extragalactic background light (EBL) attenuation \cite{2008A&A...487..837F}. Finally, for those bursts with unknown redshift, we assume two characteristic values of z$\sim$0.009 (similar to the GRB170817A \cite{2017ApJ...848L..12A}) and z=0.3 (an average value for sGRBs).

\section{Results and Conclusions} \label{Ch:Conclusions}

\begin{figure}[ht]
  \centering
    \includegraphics[width=1.0\textwidth]{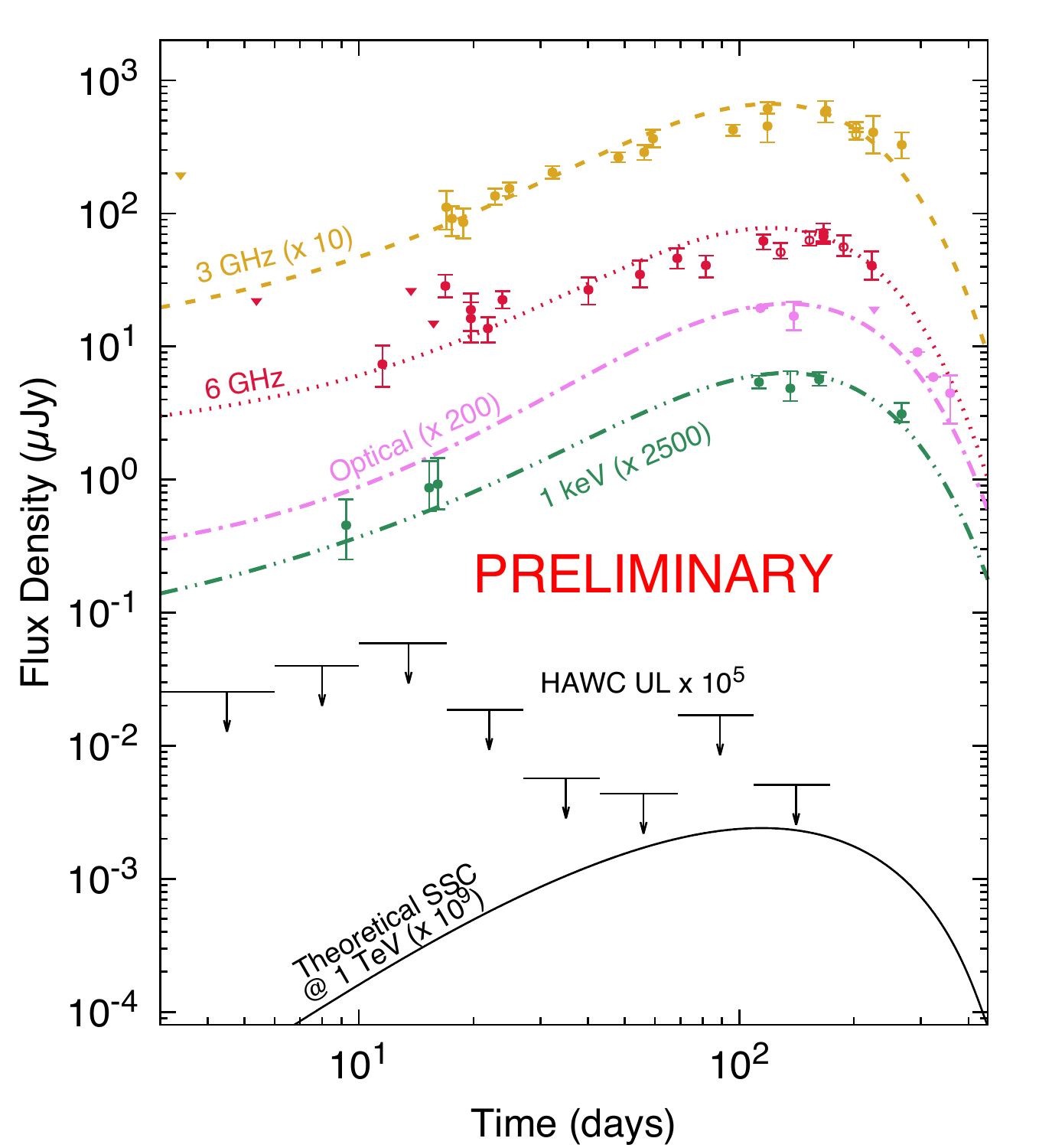}
      \caption{Upper limits derived by HAWC using 10 slide windows. The solid line is the SSC prediction at 40 TeV. \label{fig:HAWC-TimeScale-Days}}
\end{figure}

Unfortunately, there was no TeV emission found for any of the bursts studied here. In particular, the flux upper limits derived for GRB 170817A are presented in Figure \ref{fig:HAWC-TimeScale-Days}. We have estimated a theoretical prediction of SSC emission from the observed X-ray emission. The upper limits are not constraining as observed in Figure \ref{fig:HAWC-TimeScale-Days}. 

From the GRBs identified by \cite{2019arXiv190106158V}, only GRB 150101B, GRB 170111A and GRB 170817A were within the HAWC's field of view for at least one hour.
%during which HAWC recorded data at the time of their transit. 
In Table \ref{tab:SampleOfGRBs} the transit duration and the total exposure time for each of these bursts reported. 

As observed, even though GRB 150101B has the highest redshift observed, the derived flux upper limit is comparable with the one obtained for GRB 170111A assuming a redshift of 0.009. The best flux upper limit is obtained for GRB 170817A because of their closeness and medium low exposure. The flux upper limit for GRB 170111A when assuming a redshift of 0.3 is also a consequence of the very low exposure time. The decreasing behavior of the flux upper limits as a function of time is a clear consequence of the increasing time window. 

HAWC continuously monitors the whole sky in order to search for VHE emission from sGRBs in timescales from seconds to days. In this work, we present the most interesting sGRBs in the field of view of HAWC and associated or possible associated to gravitational waves.

\begin{table}[h]
\begin{tabular}{ccccccc}\hline\hline
GRB NAME    & \begin{tabular}[c]{@{}c@{}}RA\\ (deg)\end{tabular} & \begin{tabular}[c]{@{}c@{}}Dec\\ (deg)\end{tabular} & \begin{tabular}[c]{@{}c@{}}Error\\ (deg)\end{tabular} & \begin{tabular}[c]{@{}c@{}}Transit duration\\ (hr)\end{tabular} & \begin{tabular}[c]{@{}c@{}}Exposure Time\\ (hr)\end{tabular} & Redshift \\ \hline \hline
GRB 150101B & 188.0 & -11.0 & 0 & 4.51  & 464.58 & 0.134    \\
GRB 170111B & 270.9 & 63.7 & 6.7 & 0.99 & 96.467 & -        \\
GRB 170817A & 197.5 & -23.4& 0 & 2.01& 205.27 & 0.009 \\  \hline
\end{tabular}
\caption{General characteristics of the short GRBs used in this study.}
\label{tab:SampleOfGRBs}
\end{table}

%Here about figure \ref{fig:HAWC-TimeScale-Days}

%As is showed in \cite{2017ApJ...841..100A} for sources in a declination of 22$^{\circ}$

 \begin{figure}[h]
  \centering
    \includegraphics[width=1.0\textwidth]{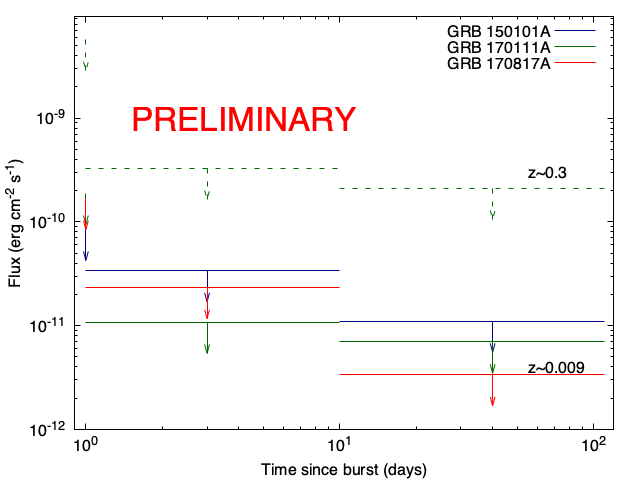}
      \caption{Upper limits for the GRB 150101B, GRB 170111B, GRB 170817A derived by HAWC into 3 temporal bins.  \label{fig:GRBs-Like-170817A}}
\end{figure}

\section{Acknowledgements}

We acknowledge the support from: the US National Science Foundation (NSF) the US Department of Energy Office of High-Energy Physics; 
the Laboratory Directed Research and Development (LDRD) program of Los Alamos National Laboratory; 
Consejo Nacional de Ciencia y Tecnolog\'{\i}a (CONACyT), M{\'e}xico (grants 271051, 232656, 260378, 179588, 239762, 254964, 271737, 258865, 243290, 132197, 281653)(C{\'a}tedras 873, 1563, 341), Laboratorio Nacional HAWC de rayos gamma; 
L'OREAL Fellowship for Women in Science 2014; 
Red HAWC, M{\'e}xico; 
DGAPA-UNAM (grants AG100317, IN111315, IN111716-3, IA102715, IN111419, IA102019, IN112218); 
VIEP-BUAP; 
PIFI 2012, 2013, PROFOCIE 2014, 2015; 
the University of Wisconsin Alumni Research Foundation; 
the Institute of Geophysics, Planetary Physics, and Signatures at Los Alamos National Laboratory; 
Polish Science Centre grant DEC-2014/13/B/ST9/945, DEC-2017/27/B/ST9/02272; 
Coordinaci{\'o}n de la Investigaci{\'o}n Cient\'{\i}fica de la Universidad Michoacana; Royal Society - Newton Advanced Fellowship 180385. Thanks to Scott Delay, Luciano D\'{\i}az and Eduardo Murrieta for technical support.

%\bibliographystyle{JHEP}
%\bibliography{bibliography_file}

\end{document}